\documentstyle[12pt]{article}
\newcommand{\f}{\begin{equation}}
\newcommand{\ff}{\end{equation}}
\begin{document}
\rightline{SU-GP-92/2-4}
\vskip .1in
\centerline{{\Large\bf Extremal variety as the foundation}}
\vskip .2in
\centerline{{\Large\bf of a cosmological quantum theory}}
\vskip .2in
\centerline{Julian Barbour}
\centerline{College Farm, South Newington, Banbury, Oxon, England}
\vskip .1in
\centerline{Lee Smolin}
\centerline{Department of Physics,Syracuse University,Syracuse, NY
13244-1130}
\vskip.2in

\centerline{ABSTRACT}
\vskip.1in
\noindent
Dynamical systems of a new kind are described.  These are
based on the extremization
of a non-local and non-additive quantity that we call the variety of a
system.
In these systems all
dynamical quantities are relational,  and particles have properties,
and can be identified, only through the values of these relational
quantities.
The variety  then measures how uniquely each of the elements
of the system can be distinguished from the others in terms of the
relational variables.   Thus a system with extremal variety is one in
which the parts are related to the whole in as distinct a way as
possible.

We study several
dynamical systems which are defined by setting the action of
the system equal to its  variety,
and find that they
have the following characterstics:  1)  The dynamics is deterministic
globally, but stochastic on the smallest scales.  2)  At an intermediate
scale structures emerge which are stable under the stochastic
perturbations at smaller scales.
3) In near to extermal configurations  one sees the emergence of both
short ranged repulsive forces and long
ranged attractive forces,
4)  The dynamics is invariant under permutations of the labels of the
particles.   For these reasons it seems possible that extremal variety
models
could provide a  foundation for a new kind of non-local hidden variables
theory, which could be applicable in a cosmological context.

In addition, the mathematical definition of variety may provide
a quantitative tool to study self-organizing systems,
because it distinguishes highly
structured, but asymmetric, configurations such as one finds in biological
systems
from both random configurations and configurations such as crystals
which are highly  ordered by virtue of having a large symmetry group.

\vfil
\eject

\section{Introduction}

This paper is addressed to the question of how to construct a physical
theory that could apply to the universe as a whole.  It is motivated
by the idea that
the problem of constructing a theory that could apply to a single
and entire
universe differs fundamentally from the problem of
constructing a theory to describe phenomena occurring
in a small portion of
the universe.   There are several reasons for this belief; perhaps
the primary one is that  a theory of a portion of the
universe can refer, in the definition of its kinematical quantities and
interpretational schemes, to things that are outside the domain of the
phenomena that are  being described.   Indeed, all of our
successful physical
theories make use of this opportunity.  Some examples of these external
elements are the fixed inertial frames of Galilean and Poincare invariant
theories, the external measuring devices of quantum mechanics and
the external  time coordinate that is the basis of all non-general
relativistic dynamics.

As long as the domain in which the theory is actually compared to
experiment involves only a portion of the universe there is no problem
with the use of these external elements.  However, when we enlarge our
ambition and attempt to construct a theory that could apply to the whole
universe, we should not be surprised to discover difficulties arising
because the familiar structures that previously were defined in terms of
things external to the system now have nothing to which to refer.  This is
the origin of  the criticisms of Newtonian mechanics known as Mach's
principle\cite{mach} and also of many of the difficulties, both
interpretational and
technical, in the study of quantum cosmology\cite{julian-qc,lee-boston}.

Thus, the problem of constructing a cosmological theory is largely the
problem of constructing a sensible theory that does not make any use of
fixed , a priori, or background structures.
Indeed, one of Einstein's main motivations
for constructing general relativity was Mach's criticism of Newtonian
dynamics exactly on this point\cite{Einstein-auto},
and he succeeded to the extent that the
only full scale example we have of  a theory which does not rely on
external or fixed nondynamical structures is general relativity itself.
Specifically, the diffeomorphism invariance of general relativity is, in a
certain sense\cite{bb2,stachel-hole}, an exact
indication of the lack of background
structure in the theory.  Thus, at this time, the problem of constructing a
cosmological
theory  consists chiefly of the problem of merging general
relativity with quantum theory, and one of the basic aspects of this
problem is the problem of constructing quantum field theories which are
diffeomorphism invariant.

There are, broadly speaking, two ways in which this problem may be
resolved.  The first is by inventing a reformulation and reinterpretation of
the quantization process---the process in which classical theories are
turned into quantum theories that is suitable to diffeomorphism invariant
theories such as general
relativity and that could stand as a theory of a single universe.  Several
such programs are currently under active development, from which  we
have learned, and will
continue to learn, a great deal that is useful.  However, we find ourselves,
for
reasons we have argued elsewhere
\cite{julian-qc,lee-boston,julian-found,oldjulianlee,lee-qc,lee-spain},
skeptical as to whether any
effort in
this direction alone can be completely successful.

The second direction to
search for a resolution to these problems is by the invention of a new
theory, based on new principles, that would reduce, in the appropriate
limits, to classical general relativity and quantum field theory.  This paper
is intended as a contribution to this second program.

The work we will describe here has been underway for a number of years,
and involves projects undertaken both singly and jointly by the authors.  In
this work we have attempted to start, so to speak, from the other side,
and to pose the question, what kind of mathematical structure could be
appropriate for the description of the physics of an entire universe?  We
have found our efforts to answer this question converging from more than
one direction on a particular kind of structure.  This structure is a certain
kind of non-local dynamical law which can be applied to a large number
of different model systems.  These dynamical systems are defined by
means of a variational principle, which is based on a type of quantity that
we call, generically, the {\it variety} of the system.  The variety is a
potential that is defined on the configuration space
of the system, i.e. it is a function of the configuration of the
system.  The variety is a non-
local and non-additive
quantity, which can only be applied to a system as a whole.  It measures,
in a certain sense, how unique, one from another,
the different parts of the system are.

We consider here two kinds of  variational principles involving the
variety.  In the first, the variety alone is extremized, in the second,
and more conventional, the variety is extremized in conjunction
with a kinetic energy term.

We find that dynamical systems based on extremising the variety
by itself have
a number of general characteristics.  1)  They are deterministic globaly,
but the behavior on the smallest scales is stochastic.   This is because the
variational
principle is not locally additive, so that global extrema are not local
extrema.  2)  The extremal configurations
have structures on an intermediate scale that are stable under the
stochastic perturbations at smaller scales.  Given a notion of
time (which we will discuss below) these structures evolve in ways that
are slow
on the scale of the fundamental time step.  3)  In spite of the overall
non-locality of the dynamics, which is
responsible for the stochastic behavior at the smallest scales, in near to
extremal configurations an approximate local dynamics emerges on these
intermediate scales.  Further, in some of the models we see the emergence of
short ranged repulsive interactions and long ranged attractive forces.
4)  The extremal configurations are, in a certain
sense, the opposite of ordered systems in that they have as little
symmetry as possible.   5)  The action principle is always invariant under
the group of permutations of the elements of the system, so that these
are deterministic dynamical systems of intrinsically identical particles.
(The precise meaning of this last statement will be clarified below.)

Because of the suggestive character of some of these results, as well as
for other reasons that we will elaborate on shortly, we
believe that it might ultimately be possible
to use this kind of dynamical
systems to
describe a  cosmological hidden variable theory.   This would be a theory
that would resolve together the problems of quantum cosmology and the
interpretational problems of quantum mechanics in a theory that was at
once a theory of cosmology and a theory of the deterministic---but
highly non-local---dynamics behind quantum mechanics.
However, we want to stress that we are not here proposing such a theory;
having come to the conclusion that a new kind of dynamical structure
would be needed in the construction of any such theory, we intend here
only to study a possible candidate for
such a dynamics.  The systems that we analyze here are thus intended only
as toy models.

Besides these long term aims, we also have the impression that extremal variety
systems are interesting in their own right.  They introduce kinds
of structures that, so far as we know, have not hitherto been studied.
Besides the
possible application to physics,  they may also be interesting for other
fields such as information theory, the study of neural networks,  the
study of biological systems or the study of self-organizing systems
in general.   In this
paper we will touch only briefly
on this aspect of the subject, it will be taken up in detail in another
place.

The directions from which we have been led to the study of extremal
variety dynamics include a search for a nonlocal dynamics that could
underlie a hidden variables theory\cite{lee-qc,lee-hidden}, a search for a
formulation
of dynamics that realizes the principles set by Leibniz in his
criticisms of Newtonian
dynamics\cite{julian-nature,bb1,bb2,julian-bj,julian-qc,lee-boston} and
the search for a dynamics
that  could be applied to a
combinatorial system of relations that would cause a condensation to a
phase that could be described in terms of the geometry of a low
dimensional spacetime\cite{lee-hidden,lee-boston}.
These motivations are recalled briefly in
the next section, but are described in detail
in the cited references.  In this paper we concentrate on describing the
models and the results of our studies of them.

In the following section we elaborate on the motivations
from quantum gravity and the interpretational problems of
quantum mechanics for attempting the construction
of non-local and relational dynamical systems.  In section 3 we give a
general characterization of extremal variety dynamical systems, and
introduce much of the language and mathematical structure that
is used to describe them.  Section 4 is devoted to examples of extremal
variety systems constructed from graphs.  These have been
described before\cite{lee-boston}, so the discussion here is brief.  Three
new
models of extremal variety systems, which are much easier to study
than the graph theory models, are described in section 5.  We close in
section 6 by summarizing what we have learned from these systems
and setting forth some conjectures and directions for future study.

\section{Motivation:  issues facing attempts to construct theories of
a whole universe}

 When one approaches the problem of
constructing a theory of cosmology, several new issues emerge which do
not have to be faced in any less ambitious undertaking.  As these issues
form the motivation for the work presented here, we will begin by briefly
sketching them.

1)   As we stated in the introduction,
all existing physical theories posit the existence of fixed background
structures which are non-dynamical, but are necessary to fix both the
forms  of the equations and the interpretive rules that describe the
evolution of the dynamical quantities that
the theory describes.  These background structures refer either explicitly
or implicitly to things that exist in the universe but are outside of the
domain of description of the theory.

A theory of the whole universe must do away with these external
background structures, for, in principle, nothing may be outside of the
domain of description of such a theory.    In terms of the question of the
nature of space and time the task of doing
away with background structures is intimately related to the
absolute/relative debate, which has been one of the central themes in the
development of dynamics from the Greeks to Einstein\cite{julian-vol1}.

A good example of a theory in which some of the background structures in
classical mechanics have been eliminated is the set of models
developed by one of
us in collaboration with Bruno Bertotti in the context of the classical
dynamics of point particles\cite{bb1,bb2,julian-nature} .
In these models a dynamics is
constructed in  which the measurements of space and time are
purely relational,  so that only relative distances play a
role, and external frames are not used.  Here one sees two key features
that emerge when one eliminates background structure in favor of
explicitly dynamical structure.  The
first is that gauge symmetries arise, which indicate that certain
mathematical structures, such as the coordinates of particles or time
parameters, are auxillary quantities and do not refer to anything that is
physically meaningful.    The diffeomorphism
invariance of general relativity is, in fact, a gauge symmetry exactly of
this kind.  Its consequence is that spacetime points are not physically
meaningful, so that no physically meaningful observables may refer to
them.

The second feature that arises, in both these model systems and in general
relativity (when cosmological boundary conditions are used), is that
certain global constraints arise which are reflected in the local physics.
Thus, in all of these cases, the
total momentum, energy and angular momentum of the universe are
constrained to be zero.  This, is of course, the only logical possibility, as
there is no external or background structure with respect to which these
quantities could be defined.  But we
believe that it is also an indication of a deeper consequence of the
elimination of background structures.  In conventional physical theories
the properties of individual particles, or of fields at individual points, are
well defined independently of
the existence of other particles or of the field values at other points,
because they are defined with respect to the background structure.
Thus, the basic entities of these theories can have a priori properties.
In a theory without background structure, this option is
not available; all
properties of individual particles
must be defined in terms of their relations with the other particles in the
universe.  This turns out to introduce new kinds of couplings between the
local
physics and the global state of the system  that are not present in
conventional background dependent physics.  Mach's principle,
in which the global distribution of matter is posited to affect the local
inertial frames, and through them the properties of individual particles,
is one example of this kind of local/global coupling that comes from
the elimination of background structure\cite{julian-nature,bb1,bb2}.

A related consequence  is that, in a theory without background
structure, there are no purely kinematical quantities.  All observable,
physically meaningful, quantities rely on dynamics for their definition, so
that the conventional distinction
between kinematics and dynamics breaks down.  This is especially
apparent in general relativity, as well as in particle mechanics with
genuine time reparametrization gauge symmetry\footnote{By which we
mean that there is no external time and the
Hamiltonian constraint is quadratic in the
canonical momenta\cite{karel-time,julian-qc}.}.  In
these
cases the problem of determining the physical observables---those
functions of  the phase space variables that are invariant under the gauge
transformations---is a dynamical problem,
and it requires the solution of the
equations of motion\cite{karel-time,carlo-time}.

We will see in the models to be described below this intertwining of local
and non-local dynamics, and of dynamics and kinematics.

2)  For the reason just cited, as well as for reasons described
elsewhere\cite{lee-hidden,lee-qc,lee-bryce}, we have
been interested for some time in the problem of
whether alternatives to quantum mechanics can be constructed that
maintain a realist ontology and epistemology.   This leads to the problem
of constructing hidden variables theories and,  through the work
of Bell\cite{bell}
and the related experimental developments\cite{aspect}, to
the problem of
constructing a non-local hidden variables theory.  This is  relevant to our
project here,  because the kind of non-locality that the experimental
results seem to call for is so drastic as to make it seem likely that any
non-local hidden variables theory must be a cosmological theory.
According to the experiments, as well as to the quantum
theory itself,  non-local correlations are established between any
two particles whenever they interact.  These correlations are carried
forever by the particles, so that at any time in the future the complete
determination of the quantum state of
a single electron requires a knowledge of the quantum states of all the
particles it has since interacted with.  If this is to be replicated by a
hidden variables theory, the hidden variables which describe a single
particle must tie it to all of the
particles it interacted with in the past.  Thus, the causal structure of
such a hidden variables theory must be a representation of the history of
the universe.

This circumstance leads naturally to the idea of relational hidden
variables theories\cite{lee-hidden,lee-qc,lee-boston}.
In such a theory the hidden variables give, not a
more detailed description of the state of an individual electron, but a
more detailed description of the
relations which exist between that electron and the other particles in the
universe.  The problem of constructing an acceptable hidden variables
theory thus leads naturally to the same issues that the problem of
constructing a cosmological theory without
background structures lead to:  how to construct a sensible dynamics in
which all quantities are relational rather than absolute quantities which
necessarily refer to a fixed background structure.  Indeed,
it was an attempt to elaborate such theories that first led to the
formulation of the idea of extremal variety\cite{lee-boston}.

3)  In conventional physical theory, the description of a given phenomenon
is divided into two parts: the specification of the laws and the
specification of the initial conditions.  As has been
stressed before\cite{karel-time,carlo-time,hartle},
this separation is somewhat problematical when we come to the case of
cosmology.    First, as the universe only happens once, or at least as we
 will only get to observe one instance of a universe, it is not clear what
empirical meaning this distinction has when applied to the
universe as a whole.  But, more importantly, what we really want and need
from a cosmological theory is a theory of origins, a theory that will
explain to us why the universe is the way it is.  If we stick to a more or
less conventional dynamical framework, then what we need is exactly a
theory of initial conditions\footnote{as has been stressed especially by
Hartle and Hawking\cite{hartle}.}.

This problem becomes especially acute if we recall that, as a result of the
time reparametrization invariance of cosmological theories,  and the
resulting blurring of the distinction between kinematics and dynamics,
there is no physical, diffeomorpism invariant, meaning to the notion of
initial conditions in the cosmological context\cite{karel-time,carlo-time}.
Both classically and
quantum mechanically in a cosmological theory there is a space of states,
but as there is no time, and no Hamiltonian, these states do not
evolve\footnote{This does not mean that the formalism cannot describe
phenomena that we
would call evolution in time.  But that time must be measured in terms of
some dynamical degree of freedom of the theory, and the observables that
describe evolution are really just describing invariant correlations
between the degree of freedom chosen  to be time and other degrees of
freedom of the theory\cite{karel-time,bryce-time,carlo-time}.
It is, indeed, in
the structure of such correlations that the dynamical content of the theory
is recorded.}.

4)  A question closely related to the question of the initial conditions is
why the observed state of the universe seems so special.  This
circumstance, which
has often been remarked upon,  manifests itself in several, apparently
independent, ways.  It is worth recalling  some of them here.

i)  As has often been noted\cite{Penrose,evolution},
the overall configuration of the Universe is very special.  It appears to
have vanishing angular momentum, its  density
(in terms of fundmental units) is
fantastically close to the critical density for a flat Friedman universe, and
the microwave background is isotropic to at least $10^{-3}$.  This is often
interpreted to mean that the universe began in a very special state,
for example, with the gravitational field completely
unexcited\cite{Penrose}.

ii)  In spite of the apparent isotropy of the microwave background and
observed galaxy distributions, the distributions of galaxies appear to
be inhomogeneous out to the largest scales that have so far been
probed\cite{galaxies}.  The most natural hypothesis that could be
made concerning
this structure is that it is the result of gravitational clumping that
occurred since decoupling.  Rather surprisingly, it seems to have
been very difficult to make a theory based on this hypothesis
work,  even with the help of dark matter.  Given that new observational
discoveries, such as the large scale streaming, and the apparent
periodic
structure continue to be made, this is a subject that may remain
open for some time.

iii)  It  appears that the values taken by a number of the
fundamental constants of nature (for example, the 17 constants of the
standard model of elementary particle physics) are special in the sense
that
were the values different by only a few percent the universe would be
vastly
different.  For example, were the neutron-proton mass difference much
bigger, there would be no stable nuclei, no stars, no nuclear physics or
chemistry\cite{evolution}.  Were it of
the opposite sign, the primary component of the
universe
would be free neutrons.  Several different examples of this circumstance
are discussed in \cite{Greenberg,Morowitz}.  For our purposes it is worth
remarking that in many of these cases it seems that the special values
the constants take allow the universe to develop more structure than
otherwise, for example, this is certainly the case with the neutron-proton
mass difference.

Until recently, modern cosmology took the view that the
large scale structure of the universe can be explained without making any
specific hypotheses as to what is actually in the universe. This may have
been plausible at a time when the simple model of a homogeneous and
isotropic
Friedmann model filled with "dust" was sufficient to account for all that
was known.  However as soon as one attempts to explain the real observed
structure of the galaxies, or understand why the overall structure of the
universe seems so fantastically special, such a view is clearly
insufficient.
In fact,  most contemporary work on cosmology, such as the dark matter
theories,
inflation and cosmic string theories, rely on some hypotheses as to the
content
of the universe.  If these do not succeed, then perhaps it will not seem too
mad to entertain the idea that there are laws that govern the large scale
structure of the universe and that these laws are intimately tied up, in the
way we discussed above, with the laws that govern local physics.

\section{Extremal variety as a dynamical principle}

To summarize the argument of the previous section, we are in search of a
new kind of dynamical theory which is characterized by the following
conditions:

1)  It is purely relational.  All physical quantities describe relations
between entities; conversely, none of the  entities posited by the theory
have any properties which are individually and independently defined;
instead all properties are described in terms of
the relations among the posited
entities.  Further, the use of fixed, background structure is to be as much
as possible minimized.

2)  It is non-local, because the experimental results  show that
the Bell inequalities are violated.

We will now describe a set of models which have these two features.  In
doing so, we will borrow some language from Leibniz's
{\it Monadology}\cite{leibniz}, as we
have found it to be the most appropriate language to use in this context.
The extent to which these models do
provide a mathematical realization of the ideas proposed in Leibniz's
writings will be discussed in another place\cite{julian-origin}.

We then posit a universe consisting of $N$ elements, which we will call
{\it monads}.  In a fundamental theory there should be  no background
space, so
the monads do not have
coordinates, or any other properties that refer to them individually.
Instead, the state of the system is
given in terms of relational quantities which are shared among monads.
More particularly, each monad has a {\it view}, which describes its
relation to the rest of the universe.

The unity of the universe finds its expression throught the requirement
that all the individual views are mutually consistent; this ensures
that the views can be regarded as all the different possible perspectives
of a unique structure that is the configuration of the universe.  Depending
on the manner in which time is to be introduced into the scheme, such
a configuration may represent either one instant of time or the entire
history of the universe.  The only a priori structure in the model is
represented by the constraints which establish how the $N$ possible
views corresponding to a given configuration do indeed fit together
self-consistently to make such a configuration.
The set of all such self-consistent
views forms the configuration space, $\cal S$, of the universe.

The question then arises as to what kind of dynamics to posit for such a
model.  As there is, at this level, no spacetime, the normal dynamical
quantities such as energy and momentum are inappropriate.    Instead, we
want to pick a dynamical principle
that will push the system into configurations where the very special
properties of the spacetime description will emerge.  The usual
dynamical quantities which are tied to spacetime symmetries---energy,
momentum and angular momentum---will then be recovered as effective
concepts.

An appropriate set of quantities which are candidates for the action
of such a non-local, relational, system,
are what we call {\it varieties}.  A variety is a
function of the views of the system  which measures how different the
different views of the system are one from another.  To
construct such a quantity we first introduce a symmetric array of {\it
differences} $D_{ij}$.   The $ij$'th is a function of the views $w_i$ and
$w_j$ and is a measure of how different they are.\footnote{In the toy
models the definition of
$D_{ij}$ will also typically include a set of symmetry operations, so that
what will be measured will be
the difference in the views mod the symmetry operations.  This has the
effect of further dimimishing the a priori structure of the model.}

The variety can then be defined as
\f
V= \sum_{i \neq j } D_{ij}
\ff
As the simplest dynamical system  we
then posit that the state of the universe is
one in which $V$ takes, at least locally, a maximal value, $N$
being held fixed.

On the face of it, such a principle will lead to the determination of one
or several (if the extremal state is not unique) static configurations.
However, this configuration could represent structure that corresponds
to both space and time.  Alternatively, time could be introduced
explicitly by introducing more conventional kinetic type terms into
the variational principle.  We shall return to this question later.

We will discuss later several alternative ways in which the variety of a
system
can be defined.  Given either the above definition (1) or
one of the alternatives,  the
variety can be understood to be, in a certain sense,
a measure of the complexity  or structure of
the system. In addition, the condition of maximal  variety can be
understood as a
condition of minimal symmetry, because if there is a symmetry operation,
$\cal S$, which is a permutation among the $N$ monads that leaves the
views fixed, then  there will be pairs of monads with  no
differences between their views.

We now go on to discuss four models in which this structure is realized.
We want to emphasize that what we are describing are models, they are
not proposals for a fundamental dynamics of the universe.  They are
designed to provide toys to investigate
the kinds of mathematical phenomena that can occur in a dynamical
system based on the principles stated above.  We believe that some  of
the phenomena we find here are suggestive of various features of the real
physical world and this, of course, encourages
us.  But for the moment this point is less important than the fact that
these models manifest a rich array of behaviors and structures.

\section{Extremal variety graphs}

The first example we will give is what might be called a pure relational
model, because
there is no background space at all.  The system is defined entirely in
terms of a
set of relational variables, one for every pair of monads.  We will describe
the simplest
example of such a system,
which is simply a graph\footnote{It seems to us that a graph is
just about the simplest mathematical concept that one could take to
represent a universe.  For it enables one to express
the idea that an interesting universe should contain genuinely distinct
entities (this part is played by the vertices of the graph) that are
nevertheless linked together in an indissoluble whole (the graph).
Moreover, the parts of the universe (the vertices) only acquire genuine
identity by virtue of their relations or connections to each
other (the role of which
is played by the lines of the graph).  In a graph of the type considered
here, all extraneous structure is pared away and we are left with what
we believe is an irreducible model of a universe\cite{julian-origin}}.

Consider, then,  a connected graph, $\cal G$, on $N$
points, such as that shown in Figure 1, to be a model of a
complete, self-contained, universe.
The $N$ points are the monads, and the state of the system is
given by the connections of the
graph.  This can be described by the adjacency matrix, $A_{ij}$, which is
defined so that, for $i \neq j$, $A_{ij}=1$ if points $i$ and $j$ are
connected and $A_{ij}=0$ if they are not connected.  We will also assume
that the diagonal elements are zero,
in accordance with the idea that monads have no a priori properties.
The state space of the system is the space of all possible graphs on N
points\footnote{It is
possible to consider generalizations in which there can be more than one
connection, or more than one kind of connection between points, however
we will discuss here only the
simplest possible model, in which each relational variable is purely
binary, the connection is either on or off.}.
It has  $2^{N(N-1)/2}$ elements, which grows
very rapidly with $N$.

We now will introduce two ways in which the view of a point of a graph
can be
defined.
The first definition of the view of a point is simply a list of other points
that it is connected to.  The {\it space of views} $\cal V$ then consists of
vectors in $R^N$ whose components are either $0$'s or $1$'s.  Thus, the
$i$'th view, $\vec{w}^i$, is given by the $i$'th
column of the adjacency matrix, so that
\f
[w_i]^j=A_{ij}
\ff
This simplest definition might be called the first order view.
(Of course, since the views are all derived from a single graph, they
must satisfy a large number of consistency conditions.)  Higher
order views can be defined which measure the points that a point is
connected to by more than one step of the lattice.  For example, one
definition of an $n$'th order view is,
\f
[w_i^n]^j=(A^n)_{ij}
\ff
The differences $D_{ij}$ are then designed to measure how differently the
$i$'th and $j$'th point are connected to the rest of the graph.
The simplest definition is based on the first order views, it is
\f
D_{ij} = | \vec{w}_i - \vec{w}_j |
\ff
where we use the Euclidean norm in $R^N$.

The variety can then be defined according to (1).

The second definition of the view of a vertex is based on a suggestion by
David Deutsch\cite{deutsch}.  In this definition, one
defines the concept of the
$m'th$
neighborhood of the $i$'th vertex.  This is defined to be the subgraph
${\cal N}^m_i$ of $\cal G$ that consists of only the vertices within $m$
steps of $i$, and the connections between them.   The original, $i$'th
point, is a preferred, or marked, point on the graph.  We then define
a notion of isomorphism between graphs with one marked point:
Two graphs are defined
to be isomorphic if there is a one to one and onto map that takes the
vertices and lines of one into the vertices and lines of the second, taking
also the marked point to the marked point.  We then define
$K_{ij} $ to be equal to the smallest $m$ such that ${\cal N}^m_i$
is {\it not} isomorphic to ${\cal N}^m_j$.   In this case it is appropriate
to call $K_{ij}$ the
{\it relative indifference} of $i$ and $j$;  for a large
value of $K_{ij}$ means that $i$ and $j$ cannot be distinguished
readily by local comparisons.
The variety is then defined in the following way.  For each vertex, $i$, we
define its {\it absolute indifference}, $R_i$,
to be the maximum value of
$K_{ij}$,
for all $j \neq i$.  This tells us how large neighborhoods we must
look at to distinguish the $i$'th point from {\it all} the others,
just by the
topological structures of its
neighborhoods.  We then define the variety to be,
\f
V^\prime =  \sum_i{1 \over  R_i}
\ff
Thus, a graph has a maximal value of the  variety if the sum of the $R_i$ is
minimized.  This means that, overall, each point is distinguished from the
others in terms of its neighborhoods by using as small neighborhoods
as possible.

This second definition is, in a sense, more intrinsic than the first,
because it makes no use of any labeling of the graph.  Using it, a vertex of a
graph is defined only by a description of its neighborhoods.

We believe that, at the most fundamental level, this is the way identity
should be conceived: all entities must be defined by
their attributes---and nothing else.  Thus, one must not think of the
vertices as little identical pre-existing points waiting waiting to be
connected up into a graph.  Rather, the vertices and lines, together with
their intrinsic identities, come into existence with the graph.

However, note that each  of
the  definitions of variety we have just given
are invariant under a gauge symmetry, under which the
vertices of the graph are arbitrarily relabeled.  One has therefore the
option of regarding the vertices as identical particles but requiring
the dynamics to be invariant under permutations among them.  This
establishes a certain similarity with the symmetry principles of
quantum mechanics that result from identity of particles; we shall
see later that fermionic behavior is to a degree inherent in maximal
variety schemes.

A preliminary study of maximal variety graphs, using the first of these
definitions, was performed several  years ago, and described briefly in
\cite{lee-boston}.    The basic difficulty is
that the very  large growth in the number of graphs
with $N$ means that statistical or  Monte Carlo methods must be used to
study graphs which are reasonably large.  Using a Vax it was possible to
study graphs consisting of up to about a thousand points.  From this
preliminary study a number of qualitative conclusions emerged:

i)  Extremal variety graphs are very special; they have much higher variety
than the average value of the ensemble of randomly generated graphs.

ii)  Random trees have higher varieties than general random graphs.

iii)  Graphs which are invariant under a symmetry operation cannot be
extremal, each symmetry operation costs some variety.  This suggests
the point of view that extremal variety graphs are as asymmetric
as possible.

Now, if a graph like this is to describe a universe, then the most important
question is
how space arises.  What we mean by space, from a purely relational point
of view, is that
the sets of relative distances among a set of $N$ particles is very highly
structured, so
that they can be represented by points in a low dimensional manifold.  In
the simplest
case where that manifold is $R^3$  this means that
the $N(N-1)/2$ relative distances are actually determined in terms of
$3N-6$ coordinates
(We subtract $6$ for the arbitrariness of the frame.)   Furthermore, if we
look at the
distribution of the $N(N-1)/2$ distances among $N$ points in some $R^m$,
we see several
characteristic features:  1)  The distribution
of distances $d$ follows  a power law, proportional to $d^{m-1}$.  2)  The
ratio of the largest
to the smallest distance is typically very large.  To see how unusual it
could be that a
system of relations like a graph could approximate $N$ points in an $R^m$,
with $N$ much
much larger than $m$, one need only appreciate that in a random graph,
such as
one defined by some probability, $p$, that any link is on,  the distribution
of
distances is Gaussian, and the ratio of the largest to the smallest distance
is, for almost
all $p$, a very slowly growing function of $N$.

We can then ask, is there a dynamics that we can posit for graphs that will
pick out the
very unusual graphs whose distance relations (determined in some
intrinsic way) could
approximate those of a distribution of points in $R^3$.  It was, indeed, as a
possible
answer to this question that the idea of a variational principle based on
the notion of
variety was invented\cite{lee-boston}.  The intuitive idea is that a graph
with a large
variety will
have a very spread out distribution of distances, so that the ratio of the
largest to the
shortest distance should grow with $N$ much faster than in a random
graph.  While behavior
of this kind was seen in the computer simulations that were done, it was
not possible to
make more stringent tests of the conjecture that an extremal variety
graph will condense
to a low dimensional space.  To do this clearly requires large scale
computer simulations, unless progress can be made in this direction
by analytic means.

\section{Some simpler models of extremal variety}

Because the graphs do not live in any background space, and in general are
hard to imbed meaningfully in any space (this is, indeed, the point) it is
often hard to see what is going on when one is working with a large graph.
Because of this, and the general computational difficulties of working
with large graphs, it is useful to define some toy models that involve
points living in some fixed background structure, but still use a principle
of extremal variety to generate structure.  This allows the general
properties of systems based on the idea of maximal variety to be studied
without having to face the particularly difficult problem of how the very
structured system of spatial relations among points in a low dimentional
manifold is generated.   More particularly, we would like to separate the
study of how a variational principle can generate structure in general
from the much more difficult question of how it can generate the
particular kinds of structure that is required if a set of points and
relations is going to condense in a way that they can be imbedded in a low
dimensional space.

It is easy to invent such systems.  We will discuss here three that have
been
studied in some detail analytically and numerically.

\subsection{A first one dimensional model}

Consider a one
dimensional lattice with $P$ sites.  The ends are identified, so that we
have a circle.  On these sites we distribute $N$  ($<P$) identical
particles (See Figure 2).
Given an appropriate definition of variety, we can then study the
properties of distributions of the particles on the lattice that extremise
it.

In such models it is appropriate to use the second definition of the
variety we introduced in the previous section.  To do this let us
introduce an arbitrary labeling $i=1,...,N$ of the particles.  A state of the
system, which consists of the positions of these particles, will be called
$\cal S$.  More properly, the state consists of the positions of the
particles mod the permutation group, as nothing will depend on the
arbitrary labeling of the particles.    For the moment, we shall allow
distributions in which more than one particle is placed on a given site.

We want to define the variety in such a way that it depends only on the
relations between the particles on the line.  We then consider the $m$'th
neighborhood, ${\cal N}^m_i$, of the $i$'th particle to consist of the $m$
sites to the left and the $m$ sites to the
right of the particle (see Figure 3).   As in the case
of the graphs, we need a definition of when two neighborhoods are
isomorphic.  To make the model more interesting, we define this in such a
way that there is no intrinsic orientation of the neighborhoods on the
lattice.  Thus, two $m$ neighborhoods, ${\cal N}^m_i$ and ${\cal N}^m_j$
are isomorphic if they are identical up to reflection around the central
point. (See Figure 4).

As before, we then define the
indifference, $K_{ij}$, to be the smallest $m$
such that ${\cal N}^m_i$ is not isomorphic to ${\cal N}^m_j$.

We may note that it may be the case for some particular state $\cal S$
that for two particles  $i$ and $j$
their neighborhoods do not become distinct before $m$ reaches its natural
upper limit $m^*$, which is $m^*=(P-1)/2$ if $P$ is odd and $P/2-1$
if $P$ is even.  This is always the case if particles $i$ and $j$ are
placed on the same site, but it also occurs for many symmetric
configurations of the particles with not more than one particle per
site. (Incidentally, analogous possibilities also exist in the graph
model.)
 We will
thus call a configuration in which there exists, for all $i $ and $j$, an $m
\leq m^*$
such that the
$m$'th neighborhood of $i$ and $j$ are distinct, Leibnizian
configurations\footnote{After Leibniz's principle of the identity of
indiscernibles.}; the remaining configurations will be called
non-Leibnizian.

We can deal with non-Leibnizian configurations in one of two ways.
We can simply say they are not allowed.  Alternatively, if for particles
$i$ and $j$ no distinguishing neighborhoods with $m \leq m^*$ exist,
we can set the indifference $K_{ij}=M$, where $M$ is any number
such that $M>m^*$.

Now, for each particle, say the $k$'th one, we may define the relative
indifference
of $k$, denoted $R_k$ to be the maximum of the $D_{ki}$, over all the $i$.
This is the order of the neighborhood it is necessary to go to distinguish
the $k$'th particle from all the others.    The variety is then defined as,
\f
V= {1 \over N } \sum_{k} {1 \over R_k}
\ff

A configuration with high variety is then one in which each particle is
distinguished from each other one with the smallest neighborhoods.

Note that if non-Leibnizian configurations are allowed among
the trial configurations the above definition of $K_{ij}$ for
indistinguishable $i$ and $j$ will ensure, especially for large $P$,
that the non-Leibnizian configurations have a variety much lower than
the extremal configurations.  In particular, this means that our
dynamical principle will automatically exclude the placing of two
particles on the same site (such configurations being automatically
non-Leibnizian).  As a consequence, the particles will exhibit fermionic
type behavior, which will be achieved dynamically and not through
imposition of a kinematical condition.

The above model was studied numerically\footnote{Using Lightspeed C on a
Machintosh II.} by means of a program which
generated states which are { \it local } extrema of $V$.  The
program begins by  generating an initial state $\cal S$ randomly.  It then
randomly picks one of the $N$ particles, say the $k$'th one, and moves
it to a
new unoccupied position, trying first one step to the left or right, then
two steps,
and so on.  It tests to see if the
move increases the variety or not, if it does it keeps it, if it does not, it
goes on to try the next unoccupied spaces to the left or right.
The program stops when there is no move for any of the particles
that will increase the variety of the state.  The resulting state,
${\cal S}_{ext}$ is then a local extremum of $V$.

Some
local extremum of $ V$ are shown in Figures 5-8 for  several different
$N$'s and $P$'s.    From these results we may draw some
conclusions:

1)  The maximal variety states are highly structured.  It is, indeed, easy to
distinguish each point from all the others by inspection, by noting visually
its neighborhood.

A maximal variety state is then very different from both a random state
and a highly ordered state.  We will discuss this distinction further below.

2)  Even for these relatively small $N$ and $P$, there are several distinct
local extrema.   This tells us that $ V$ must be a rather complicated
function on the space of states.  For example, in Figures 6-8  we see
several different local extrema for the same $N$ and $P$.  These
configurations are similar, in the qualitative appearance, but distinct.

For example for small $N$ compared to $P$ there are two or more
independent groups, separated by quite a bit of space. The tendency of the
particles to form groups is a result of the fact that if there is more than
one lone particle, each separated by quite a bit of space, it will take a
large $m$ to distinguish them.  This $m$ can be decreased if at least one
of them moves close to a group that can distinguish it.  Thus, as long as
there is more than one lone particle there is what can be thought of as a
long range attractive force between lone particles and groups and among
lone particles.  (Indeed, with the form (6) of $V$ the potential pulling lone
particles to groups is  proportional to the inverse distance!)   In a solution
the variety is extremized, and there cannot be a situation with more than
one lone particle.   However, there can   be left out exactly one lone
particle, as this will be easily distinguished from all the others.  Indeed,
from
the results we see that in the extremal configurations there is often one,
but never
more than one, lone particle.

At the same time there must also be a repulsive short ranged interaction,
because to extremize the variety the groups should all be distinct from
each other.  We cannot have too many identical groups of two or three,
each with a lot of space around them.

The emergence of both long ranged attractive forces and short ranged
repulsive forces are examples of
how local dynamics can emerge from a global, non-local variational
priniciple.    We note, in particular, that the short ranged repulsive
forces are very similar in their action to the degeneracy pressure
in fermions that results from the kinematic Pauli principle.  (We have
already shown how the requirement of extremal variety effectively
prevents the occurrence of configurations with more than one particle
at a given site).

To see how the structure of $V$ on the configuration space exhibits both
these short
ranged repulsive forces and long ranged attractive forces,  we may
consider a
sampling of configurations through which the system approaches the
extremum.  This is indicated in Figures 9-11 .  In these Figures we show
several
numerical experiments in which we begin with a random configuration
with $N$ particles and watch how they move towards an extremal
configuration.   We see that, characteristically, the   particles tend to
move  by sequences of steps which are occasionally punctuated by large
jumps.    The overall tendency is for the particles to pull together into
groups, exactly as if they are moving under the influence of attractive
forces.

It is interesting to note that  if the new particle is added
to a position far from any groups, and the original configuration already
has a lone particle, both lone particles will move towards the nearest
groups, which then absorb them.  If, however, the new particle is added to
a point in the midst of a group, it will likely have a small, but nonlocal
and apparently random influence on many of the other particles as they
shift around to find the configuration in which all are best distinguished.

We see that in both cases, the number and center of masses of the large
groups are not changed by the perturbation of adding a new particle, even
in the case that the small scale configuration of each group shifts quite a
lot.  This is an example of a further feature of the solutions which is of
interest, namely at the intermediate
scales of the groups, the structure of the
solutions is stable against small perturbations.  At the same time these
perturbations cause changes at the smallest scale  that are apparently
random (in the sense that they have no local cause).

We can in addition say some things about the groups
that the particles form
themselves into.  First, as we already remarked, each of these groups is
different from all
the others.  This is necessary, of course, so that each member of each
group is distinguished from all the others by as small neighborhoods as
possible.    If we look only at one particular group, consisting of a certain
number of particles,  we cannot predict exactly what form it will take.
This illustrates a second feature of the solutions that we find very
interesting, that the behavior of the system on small scales appears
random if one looks for explanation in terms of a local law.  At the same,
time, if one can look at the whole system one finds that the exact
configuration is a solution to a non-local variational principle.

Finally, we see that because in an extremum there will be at most one
group consisting of just one particle, the long range forces have
disappeared, and there is no change in the variety if the overall distances
between the groups change, as long as they remain large compared to the
size of the groups.    As we remarked above, in an extremum the long range
interactions must disappear because every particle will be
distinguished from every other one by small neighborhoods.

It is also interesting to look at the sequence of extrema which are
produced by these
experiments.  These are shown in Figure  12.  Each
of these shows a sequence of extrema, each of which has been produced
from the previous one by adding a monad at a random point and then finding
a nearby
extremum.   (Thus, each line in these Figures corresponds to a line before a
space in
Figures 9-11.)   We see again the combination of local
stochasticity together with stability of structure on larger scales, as each
of the
groups grows, or new groups are started.  This
continues until  the groups come close enough to each other that the short
range interactions which attempt to keep the groups distinct begin to
involve the particles in different groups, at which point they effectively
merge.

We now turn to the discussion of a similar, but not identical, one
dimensional model.

\subsection{A second one dimensional model}

In the first model we have put in a large amount of background structures
in order to bring out in an easily visualized way how a variational
principle based on variety produces structured configurations.   However,
as the basic conjectures which motivated the introduction of this kind of
variational principle are based on the idea that we seek a theory with as
little background structure as possible, it is interesting to ask to what
extent we can eliminate the background structure in this model and still
have a system which is easy to study.  Of course, the basic structure we
have put into the model is the one dimensional lattice and we do not want
to eliminate this, or we are back with the much more difficult problem of how
space arises.  However,
there are other aspects of background structure which are put into the
model.  For example, there are two kinds of sites on the lattice, occupied
and unoccupied.  In the model these are treated asymmetrically in the
definition of the variety.  Further, if we look at the solutions we see
that there is no symmetry between the occupied and unoccupied sites.

Now, we might imagine a model in which this distinction between
occupied and unoccupied sites is not put in at the beginning, so that the
dynamics is completely symmetric under the exchange of the occupied for
the unoccupied sites.  It is then interesting to ask whether the solutions
will break the symmetry.   If this happens, then we will have a further
example of the way in which a variational principle of the sort we have
been discussing can introduce structure from an a-priori symmetric
situation.

Our second one dimensional model is thus constructed so as to be
symmetric in occupied and unoccupied sites.   In this model we consider,
again,  a closed one-dimensional lattice with $P$ sites.  On each slot we
put either a black ball or a white ball, each such
configuration constitutes
a state $\cal S$. If $W$   is  the number of white balls,  and $B$ is the
number of black balls, we have, because each site is filled,
$W+B=P$   For each
pair of distinct sites $i,j$  we form the {\it relative indifference}
$K_{ij}$  as follows: We attempt to distinguish positions $i$ and $j$ in a
way that is symmetric under both the interchange of the labels, white and
black, and the exchange of left and right. For example, we might say that
both neighbours of $i$  have the opposite colour to $i$,  whereas one of
$j$'s  immediate neighbors has the same colour as $j$. Under such
circumstances, $i$ and $j$ can be said to be  1-step distinguishable and
we define $K_{ij}$ to be 1. However, it might be necessary to extend the
comparison of the neighbourhoods of $i$ and $j$ to $m$ steps before the
two positions can be distinguished; then $K_{ij}=m$. In making the
comparison of the $m$'th neighbourhoods, we again use a definition in
which two neighborhoods are identical if they are the same up to the
exhange of black and white and left and right (See Figure 13).

Non-Leibnizian configurations can be treated as in the previous models.

The complete set of relative indifferences $K_{ij}$ again forms a
symmetric matrix (whose diagonal elements we set equal to zero by
definition). The variety is then defined from this matrix.  For convenience,
in this model we work with the inverse of the variety, which we call the
{\it total indifference}.  This may be defined  in two different ways:
either by finding the maximum value of $K_{ij}$  in each row $i = 1,2, .., N
$and adding these $N$ largest indifferences to give a total indifference
(this is analogous to the definition we used in the previous model) or by
simply adding all the $N(N - 1)/2$ relative indifferences.

We have found in our calculations that the actual definition we used for
the variety does not change the qualitative results, although the detailed
form of the extremal solutions are, of course different. In our study of
this second model we found it particularly useful to use this second
definition, so that,
\f
V^{\prime \prime}={1 \over \sum_{i <j} K_{ij}}
\ff

In our study of this second model, we were particularly interested in
properties of the distribution of varieties over the states, such as  the
occurrence of both absolute and relative extrema.   This particular
definition is particularly well suited to this study, because the variety is
the inverse of an integer.  For example,  with such a  definition it follows
trivially  that there must be a configuration of largest  variety.  Since the
inverse of the variety (the absolute indifference) is a positive integer,
there will, for each $P$,  be a least integer which is the inverse of the
variety.   Using, again, a Macintosh, we found  it was not difficult to
construct all possible configurations having a given number of slots up to
$P=25$ and study both the distribution of the variety over the ensemble
and properties of both the absolute and relative extrema.

 We now describe the results we obtained for all $P \leq 25$.   First, for
$P < 7$ there are no configurations in which every slot is distinguished
from every one just by a description of its neighborhoods, mod the
symmetries of the model. These are then all non-Leibnizian.   For $P= 7$,
there is only  one Leibnizian configuration, which takes the form
\f
xx-x---,
\ff
Here, we have chosen arbitrarily that the $x$'s represent one color, say
black, while the $-$'s represent the other.  In this and the configurations
below, we have also   chosen a representation in which the largest string of
$-$'s is to the right; because of the periodic boundary conditions and the
left-right symmetry this is purely arbitrary, but it helps to compare
results for different $P$.   Thus,  the following configurations are
identical descriptions of the state depicted in (8)
\f
--x-xxx    \ \ \ \ \
x-xx---
\ff

As $P$ is increased, the number of distinct Leibnizian configurations
increases quite rapidly, it is of the order of 1.2 million for $P=27$.
For each $P$, up till 25, we found all configurations which maximized $V$.
These are
shown in Figure (14).

The first striking thing about these results is that for most $P$ there is
not a single global extremum of the variety.  Instead there tend to be a
number of distinct configurations which realize a single  extremal value.
(Because   $1/V$ is an integer, this is possible.)  The  number of such
configurations is not large; and does not seem to be increasing, for
example, for $P=14$  are nine, for $p=15$  there are three, for $P=22$
there are four.  In addition,  the extreme value of $1/V$ is often realized
by configurations with different numbers of black and white balls.
However, the occurrence of these multiple extrema is, perhaps, not
puzzling in light of the rather large numbers around.   For all $P$ except
the first few, the number of Leibnizian configurations is much greater
than the maximal value of $1/V$, thus for each integer that $1/V$ takes
there will be many configurations.

As in the case of the first model, the maximal variety configurations
up to $P=25$ all possess certain characteristic features. First, we
see that the symmetry between black and white is spontaneously broken
by the solutions, there is, in fact, no solution which maintains the
symmetry of the dynamics.  Indeed, the symmetry is broken in both the
number and position; in all the extremal configurations with $P > 14$,
significantly less
than half the slots are of one color.  Second,  about one third of
each extremal
configuration is occupied by a uniform run of sites of all the same type.
As represented above, these are the strings of zeros on the right of all
configurations. We shall call this uniform run, perhaps wistfully, {\it the
space}.   In all cases, the space is bounded at one end by a single site of
the opposite type followed by another site of the same type as the space.
At the other end of the space, there are always two or three sites of the
opposite type. After that the two types of site seem to alternate in a
manner that is very hard to predict; without doing the exact calculation it
seems to be impossible to say what will be found in the part of the
configuration that is not the space.

It is interesting to examine the matrix of relative indifferences, for one of
these
configurations.  This is shown, for one of the $P=25$ configurations, in
Figure 15.

It is perhaps worth commenting here on a characteristic feature of the
maximal variety models that makes them very different from conventional
dynamical systems.  These are usually based on variational priniciples in
which the action is a local additive quantity; as a consequence, any
subsystem of a large system that is in an extremal state is itself
also in an extremal state.  This is not true in maximal
variety schemes because of the extreme nonlocality of the variational
principle.  Thus, {\it parts} of the configurations shown in Figure (14)
are very far from being in a state of maximal variety, even though the
complete configurations are extremal.  Indeed, what we have called
the space is entirely devoid of variety.  Intuitively, it is clear that
interesting structure must have a background of less interesting
structure to set it off.  In maximal variety configurations both the
background and the structure have been generated dynamically.

To conclude we see that the results of the study of this second model
reinforce the results
from the first model.  We may then summarize the features of the
extremal
configurations of the one dimensional models as follows:
1)  On an intermediate scale there develop stable
structures, while at the same time the small scale structure is apparently
random.  These structures consist of groups of one to four  particles
organized in a distinct way.  2)  These structures are apparently stable
under the
small scale perturbations that arise when an additional particle is added.
3)  The extremal configurations are intricately organized, but have no
apparent
symmetries.  4)  We gain some information about the form of $V$ in
neighborhoods of
extremal configurations.  In particular, we see evidence that lone particles
feel long ranged attractive forces to groups and to other lone particles, as
long as there is more than one lone particle in the configuration.  We also
see evidence
of short ranged repulsive forces, which are necessary to keep the groups
distinct.
5)  Finally, the fact that rather similar structures arise in the two models,
which use
different definitions of both the kinematical state and the definition of
the variety,
suggests that these characteristics are rather general, and depend only
weakly on
the specifics of the definition of the variety that is being used.  This
reinforces the
idea that these definitions do capture the intuitive idea of variety of
views that motivated
the mathematical definitions.

Before going on to the next model, we should remark that the two
one dimensional models we have described here can be generalized to any
higher
dimension.  This has not, so far, been done.

\subsection{A two dimensional model}

Another kind of model for maximal variety, also involving a distribution of
points in a fixed space can be constructed as follows.  Consider a
scattering
of $N$ points, representing our monads, in the two dimensional plane;
their positions are given as
usual by the points $\vec{x}_i$,  $i=1,...,N$.  We can define a variational
principle literaly in terms of their views, i.e, in terms of what they "see"
when they look around them.  What the $i$'th monad sees is a distribution
of
points on the circle, which can be written $\rho (\theta)_i$.  We then can
define the  indifference matrix, $K_{ij}$, between two monads to be a
measure
of the difference between their two distributions. As we do not want
any preferred directions in the model, we can build into the definition a
gauge symmetry in which each distribution can be rotated by an arbitrary
angle, i.e.
\f
\rho(\theta )_i \rightarrow \rho(\theta +\alpha_i )_i  .
\ff
Since each $\alpha_i$ is independent, the gauge symmetry is $U(1)^N$. The
variety may then be defined according to one of the two schemes discussed
in section 4.

A model of this kind has been studied numerically by
Nick Benton.  In
this model,
which will be described in more detail in a separate
publication\cite{withnick}, the
circle
is divided into octants and the distribution $\rho(\theta )_i$
is described by a histogram
 of the number of points that appear in each octant from the $i$'th point.
 The difference matrix is defined to be the minimum number of points by
which the
 histograms differ, under the group of rotations by multiples of $\pi/4$.
 The variety is then defined by equation (6).

 Some first results of this study are shown in figures (16)-(19).  In Figure
(16) we
 see a distribution of the varieties in an ensemble of random
distributions of $50$ points in the
 plane.  In Figures (17)-(19) we see three configurations of very high
variety.
 They are not actually extremals, but from Figure (16) we can see that they
have large variety compared to the random ensemble  (the small arrow to the
right indicates where the configurations in Figures (17)-(19) are
in the distribution.)  We see again the characteristic
features of extremal configurations that we found in one dimensional
models. The distributions are highly asymmteric and highly structured, with
different kinds of structure appearing on different scales.  At
the smallest scales,
many but not all of the points are clumped in groups of two to four.
These clumps are then distributed with respect to each other in a way
that is apparently ordered on large scales; by the eye it appears that this one
dimensional structure involves the clumps wanting to line up to form
one dimensional structures.  To what extent this one dimensional ordering
 of the clumps is real, or is an artifact of the eye's tendency to put order
in random distributions is presently under investigation.

\section{Remarks}

Before closing we would like to make three remarks concerning possible
applications of the  maximal variety models.

\subsection{Order and structure.}

It is clear intuitively that there is a large and qualitative difference
between what we might call an ordered configuration and a structured
configuration.  By an ordered configuration we mean a configuration of
high symmetry, like a crystal.  By a structured configuration we mean
something like a living cell, a DNA molecule, or a large modern city
in which all of the parts and their configurations are highly interrelated,
but where there is no overall symmetry.  It is clear that both types
of configuration are far from random, highly ordered and highly structured
configurations each have very small measure in random ensembles.  Thus,
each are characterized by low measures of entropy.  However, it is also
clear that ordered and
structured
configurations are very different from each other.  A simple measure of
the
lack of randomness in a configuration, such as entropy,
does not capture this difference.  Both a crystal and a neural
network have very low entropy.

This situation seems responsible for a certain amount of confusion,
because while we
have a strong intuitive feeling as to the difference between order and
structure, between a piece of ice and a brain, it has not been clear how to
make this intuition quantitative.  It appears to us that variety, as
we have defined it in this paper, does distinguish cleanly between order
and
structure, and could thus be useful for the study of the systems in which
structure is generated.

Variety does this because configurations of high symmetry have very low
variety.  On the
other hand, configurations with high variety appear to be highly organized
in the sense that cities and biological systems are---the
parts are distinct,
but in ways that are highly related to each other.  Further,
a certain amount of small scale randomness coexists with an
organization that can only be perceived by studying the system globally.
On the other hand, random configuations come in the middle.  If we
consider
Figure (16), which shows the distributions of varieties in a random
ensemble
we see a Gaussian distribution which is peaked at an intermediate value
of the variety.  Highly ordered systems are far to the left of the random
distribution; they have very small varieties.  Highly structured systems
are far to the right; they have large varieties.

To summarize this point, varieties, as we have defined them here,
are thus a collection of quantities that
distinguish between the three kinds of configurations, ordered, random
and
structured.  As such they could take us a step nearer to the goal, which has
hitherto proved so elusive, of finding a rigorous formulation of the intuition
that brains are different from crystals,
as symphonies are different from a
busy signal.

In this connection, we should like to mention Chaitin's interesting
attempt \cite{chaitin}
to capture mathematically the essence of structured complexity by means
of an algorithmic
version of Shannon's concept of {\it mutual information}.  Without going
into details,
we merely mention that the most important part of Chaitin's approach
is the identification
within a complete structure of substructures that are isomorphic to
each other.  This is
also central to our approach, since the isomorphic (and non-isomorphic)
neighborhoods
that we use to define variety are substructures in Chaitin's sense.
Thus, at the level
of kinematic characterization, mutual information and variety
represent two different ways
of measuring structured complexity of a complete system in
terms of isomorphisms
between its parts.  Chaitin's approach is more general, but
ours may be attractive
in the prospect it offers for establishing organic connections
between algorthmic
complexity and conventional dynamics.  The fact is that
{\it identity and identification},
prime concerns of epistemology and information theory,
{\it have dynamical potential}\footnote{See also \cite{julian-origin}.}.

\subsection{Two ideas about time}

The basic theme of this paper is that maximal variety models might
provide
the mathematical foundations for new kinds of laws of physics that are
intrinsically cosmological.  Apart from the motivations described in
the first two sections, we would like to keep away from any detailed
speculations about how such laws are to be constructed.  It is far too
soon for such speculation, a great deal more needs to be known about these
kinds of dynamical systems before such ideas can be seriously entertained.
However, one aspect of the problem should be addressed: how is
time to fit into such a dynamical scheme?  We would like to suggest here
two possible ways that might be further explored\footnote{A third possibility,
that a single maximal variety configuration could represent structure
in both space and time, has already been mentioned above, and is
elaborated in more detail in \cite{julian-found}.  See also the second
reference of \cite{karel-time}.}.

\subsection*{Variety as potential energy}

As we have already noted, it
is rather natural to think of variety in the systems we have described
as a kind of non-local potential energy function.  We have seen that while
in random configurations the variety is a completely non-local function, in
near to extremal variety configurations we see evidence of both long
ranged attractive forces, which hold the groups together, and short ranged
forces, which insure that they are distinct from each other.
We have noted the interesting similarity between this latter behavior and
the consequences of the kinematic Pauli principle for fermions.  It would
then be very interesting to study  a dynamical system, say for a system
of particles in $R^n$, generated by a
conventional form of a variational principle,
\f
S= \int dt  ( T + V  )
\ff
where $V$ is the variety of the system and $T$ is a conventional kinetic
energy.

There is a further direction for the development of such models, which is
natural in view of their motivations in terms of the elimination of
background structures.  This is to combine the notion of extremal variety
with the Machian dynamical systems developed by one of us in
collaboration
with Bertotti\cite{julian-nature,bb1,bb2}.  In
this case the variational principle would be of the form,
\f
S= \int dt \sqrt{T^\prime} \sqrt{V}
\ff
where $V$ is again a measure of the variety and $T^\prime$ is now a
kinetic
energy measured in terms of the intrinsic derivative\cite{bb2}.
The product form now guarantees that the dynamics is invariant under
time reparamaterization; the intrinsic
derivative is defined independently of any background inertial
frame.  This form also
lends itself very naturally to a situation where the individual particles
are intrinsically identical, so that the label of the particles can be
permuted by a time dependent gauge transformation.

We are presently studying such systems.  We conjecture that there are
systems
of this form which can serve as non-local hidden variable theories for
quantum mechanics, along the lines of the theories proposed by one
of us in\cite{lee-hidden,lee-qc}. Further developments in this direction
will be
reported if
they turn out to be fruitful.

\subsection*{Time as creation}

 A more radical notion of how time could be described in terms of extremal
 variety dynamics is the following.  Let us suppose that at each time
 the universe is in an extremal variety configuration for $N$ monads,
 and that time corresponds to the successive addition of monads to the
system.   In Figure (12) we see the result of stacking successive
local extrema, each generated from the previous  one
 by the addition of one particle.  We see that independent groups appear
 which evolve in a way which seems slow and, to a good approximation,
 deterministic, on an intermediate scale, while the evolution at the
 smallest structure is stochastic.  Indeed, this is exactly the kind
of behaviour that we observe in the
world around us, in which the gross structure seems to evolve in
accordance with deterministic laws of classical physics, while the
microscopic structure seems to obey probabilistic quantum laws.

\section{Conclusions}

In this paper we have explored the consequences of an argument, the main
points of which are 1) A theory of the whole universe must differ
substantially
in its basic kinematical and dynamical structure from the theories that
have
so far been constructed in physics.  2)  Such a theory must be relational;
all kinematical quantities must refer to relations between the
fundamental
entities; no particle, or field,
should have any a priori properties
that are determined independently of
whether or not anything else exists in the universe.  3)
Dynamical
systems in which the variables are relational are not very familiar to us;
they
are intrinsically non-local and may be non-additive, and thus
cannot be expected
to
behave like the usual local systems we are familair with.

Motivated by these considerations, we have invented and explored a new
kind of
relational dynamical system, which is based on the notion of extremal
variety.
These systems are based on an action principle, but the quantity which is
extremized is proportional to a measure of
the differences between the way each pair of
entities is related to the rest of the system.  This quantity is,
as we have tried to
argue,
a measure of how structured the system is.

In the four model
systems we have studied so far we find several
common features that, at a rather
general level, encourage our belief that dynamical systems based on
extremal variety  deserve more study as possible new approaches to
combining
quantum theory with a theory of space, time and cosmology.  These
features
are:  1)  The dynamics is deterministic globally and non-locally, but
stochastic
on the smallest scales.  2)  Symmetry under exchanges of the labels of the
particles is built in from the beginning, so that the particles which are
described are intrinsically identical.  They are distinguished only by the
values of the relational dynamical variables.  Both of these are
features  we would  like a hidden variables
theory for quantum mechanics to have.   In addition, there is
a hint that fermionic behavior can be recovered
directly from the basic variational principle and may
not need to be imposed a priori as a kinematic condition.
3)  In extremal and near to extremal
configurations the particles tend to clump in groups that are stable under
perturbations and (under at least one notion of time---the second described
above)
evolve slowly and smoothly on the fundamental time scale.  While too much
cannot
be claimed from the results we have so far, let us say only that this
behavior
is reminiscent of how we would like a classical world to emerge from a
fundamental theory.  4)  By looking at near to extremal configurations, as
well as how the systems approach local extrema, we can learn some
general properties of the variety, as a function on the configuration space.
If we consider $V$ as analogous to a potential energy, then we see that
in near to extremal configurations it is characterized by  the
presence of long ranged attractive forces and short ranged repulsive
forces.  Again, it is premature
to claim too much from this, but it is suggestive.

We would like to close by stressing that all of the results we have reached
are preliminary; they are the result of simple computer simulations done
on
simple model systems.
Much more sophisticated studies need to be done, both numerically and
analytically, before we can have a clear idea of whether or not extremal
variety systems are the right answer to the problems posed in the
introduction
to this paper.

\section*{Acknowledgements}

We would like to thank David
Deustch for suggesting the second
definition of variety defined in section 4, and
we would like to thank Nick Benton for
permission to show the results of his computer simulations (Figures (16-19))
here
in advance of the publication of that work.  We thank the two of them and
Bruno Bertotti,
James Hartle, Chris Isham, Louis Kauffman, Karel Kuchar and Carlo Rovelli
for discussions concerning these ideas.  Hospitality by Syracuse and Yale
Universities and by the Barbour family was essential to the carrying out of
this work.  Finally,
this work was supported by grants from the National Science
Foundation to Syracuse
and Yale Universities.

\end{document}